\documentclass[fleqn,usenatbib]{mnras}

\usepackage[T1]{fontenc}

\DeclareRobustCommand{\VAN}[3]{#2}
\let\VANthebibliography\thebibliography
\def\thebibliography{\DeclareRobustCommand{\VAN}[3]{##3}\VANthebibliography}

\usepackage{graphicx}	% Including figure files
\usepackage{amsmath}	% Advanced maths commands
\usepackage{amssymb}	% Extra maths symbols
\usepackage{newtxtext,newtxmath}
\title[Constraining 21cm ionised bubbles around LAEs with MWA]{Constraining the 21~cm brightness temperature of the IGM at $z$=6.6 around LAEs with the Murchison Widefield Array}

\author[C.~M. Trott et al.]{Cathryn M. Trott,$^{1,2}$\thanks{E-mail: cathryn.trott@curtin.edu.au}
C.~H. Jordan,$^{1,2}$
J.~L.~B. Line,$^{1,2}$
C.~R. Lynch,$^{1,2}$
S. Yoshiura,$^{5,7}$
B. McKinley,$^{1,2}$
P. Dayal,$^3$
\newauthor
B. Pindor,$^{5,2}$
A. Hutter,$^3$
K. Takahashi,$^{6}$
R.~B. Wayth,$^{1,2}$
N.~Barry,$^{5,2}$
A.~Beardsley,$^8$
J.~Bowman,$^8$
R.~Byrne,$^9$
\newauthor
A.~Chokshi,$^{5,2}$
B.~Greig,$^{5,2}$
K.~Hasegawa,$^{13}$
B.~J.~Hazelton,$^9$
E.~Howard,$^{11}$
D.~Jacobs,$^8$
M.~Kolopanis,$^8$
\newauthor
D.~A.~Mitchell,$^{2,11}$
M.~F.~Morales,$^9$
S.~Murray,$^8$
J.~C.~Pober,$^{10}$
M.~Rahimi,$^{5,2}$
S.~J.~Tingay,$^1$
R.~L.~Webster,$^{5,2}$
\newauthor
M.~Wilensky,$^9$
J.~S.~B.~Wyithe,$^{5,2}$
Q.~Zheng$^{12}$
\\
$^1$International Centre for Radio Astronomy Research (ICRAR), Curtin University, Bentley WA, Australia\\
$^2$ARC Centre of Excellence for All Sky Astrophysics in 3 Dimensions (ASTRO 3D), Bentley, WA, Australia\\
$^{3}$Kapteyn Astronomical Institute, University of Groningen, P.O. Box 800, 9700 AV Groningen, The Netherlands\\
$^4$Faculty of Advanced Science and Technology, Kumamoto University, Japan\\
$^{5}$School of Physics, The University of Melbourne, Parkville, VIC 3010, Australia\\
$^{6}$International Research Organization for Advanced Science and Technology, Kumamoto University, Japan\\
$^{7}$Mizusawa VLBI Observatory, National Astronomical Observatory Japan, 2-21-1 Osawa, Mitaka, Tokyo 181-8588, Japan\\
$^{8}$Arizona State University, Tempe, AZ, USA\\
$^9$Department of Physics, University of Washington, Seattle, WA 98195, USA\\
$^{10}$Brown University, Department of Physics, Providence, RI 02912, USA\\
$^{11}$CSIRO Astronomy and Space Science (CASS), PO Box 76, Epping, NSW 1710, Australia\\
$^{12}$Shanghai Astronomical Observatory, China\\
$^{13}$Graduate School of Science, Nagoya University, Japan\\
$^\dagger$ARC Future Fellow%
}

\date{Accepted XXX. Received YYY; in original form ZZZ}

\pubyear{2021}

\begin{document}
\label{firstpage}
\pagerange{\pageref{firstpage}--\pageref{lastpage}}
\maketitle

\begin{abstract}
The locations of Ly-$\alpha$ emitting galaxies (LAEs) at the end of the Epoch of Reionisation (EoR) are expected to correlate with regions of ionised hydrogen, traced by the redshifted 21~cm hyperfine line. Mapping the neutral hydrogen around regions with detected and localised LAEs offers an avenue to constrain the brightness temperature of the Universe within the EoR by providing an expectation for the spatial distribution of the gas, thereby providing prior information unavailable to power spectrum measurements. We use a test set of 12 hours of observations from the Murchison Widefield Array (MWA) in extended array configuration, to constrain the neutral hydrogen signature of 58 LAEs, detected with the Subaru Hypersuprime Cam in the \textit{Silverrush} survey, centred on $z$=6.58. We assume that detectable emitters reside in the centre of ionised HII bubbles during the end of reionization, and predict the redshifted neutral hydrogen signal corresponding to the remaining neutral regions using a set of different ionised bubble radii. A prewhitening matched filter detector is introduced to assess detectability. We demonstrate the ability to detect, or place limits upon, the amplitude of brightness temperature fluctuations, and the characteristic HII bubble size. With our limited data, we constrain the brightness temperature of neutral hydrogen to $\Delta{\rm T}_B<$30 mK ($<$200 mK) at 95\% (99\%) confidence for lognormally-distributed bubbles of radii, $R_B =$ 15$\pm$2$h^{-1}$cMpc.
\end{abstract}


%%%%%%%%%%%%%%%%%%%%%%%%%%%%%%%%%%%%%%%%%%%%%%%%%%

%%%%%%%%%%%%%%%%% BODY OF PAPER %%%%%%%%%%%%%%%%%%

\section{Introduction}
Lyman-$\alpha$ emitting galaxies (LAEs) are strong probes of the ionisation state of the intergalactic medium \citep[IGM,][]{furlanetto06,hutter14,ouchi10,ouchi20}. Rest-frame Ly-$\alpha$ is a strong UV emission line (1216 Angstrom) produced in star-forming regions, which is readily absorbed by resonant neutral hydrogen gas due to its large optical depth. During reionisation, the optical depth to Ly-$\alpha$ emission from Lyman-Break Galaxies steadily increases in redshift between $z$ = 6--7, suggestive of a corresponding increase in neutral fraction \citep[][and references therein,]{hoag19}. Strong LAEs at the tail end of reionisation are good tracers of ionised regions, where overdensities of galaxies have excavated their local IGM such that Ly-$\alpha$ is redshifted out of resonance before encountering the partially-neutral IGM. The Ly-$\alpha$ damping wing has been used to study absorption by the co-located IGM and constrain the neutral fraction at the end of reionisation \citep{rhoads01,santos04,mesinger08}, and Ly-$\alpha$ opacity along sightlines for lower redshift sources have been used to constrain the bubble size \citep{bosman18}. The Lyman Alpha Galaxies in the Epoch of Reionization (LAGER) survey in the COSMOS field for LAEs observed an evolution in the LAE luminosity function between $z$=6.9--6.6, consistent with a model where ionised regions surround the LAE galaxies \citep{zheng17}. This hypothesis is supported by hydrodynamic and radiative transfer simulations, which show a strong anti-correlation of 21~cm emission from the IGM and Ly-$\alpha$ emission from galaxies on scales of 5$-$50 cMpc \citep[2$-$10 arcmin at $z$=6.6;][]{hutter17}, and the first observational evidence for ionized regions created by galaxy overdensities through the Ly-$\alpha$ emission \citep{castellano16}. There is thus potential for using the spatial locations of known LAEs as tracers of ionised regions in redshifted 21~cm data.

More recently, simulations of the 21~cm field around early galaxies has illuminated the topology of reionisation \citep{elbers18,chen19,kakiichi16,hutter21,sobacchi16,pagano21}, and the evolution of the size and distribution of ionised regions around UV-bright sources \citep{lin16}. \citet{heneka17} used radiative transfer simulations to map the diffuse and galactic Ly-$\alpha$ emission to 21~cm maps, showing a good agreement on smoothing scales of a few megaparsecs, confirming the anti-correlation of earlier work \citep[e.g.][]{jensen12}. Simulation studies suggest a log-normal bubble size distribution \citep{lin16,furlanetto05}, but the characteristic size is uncertain and depends on several factors including the timing of reionization. After bubbles have percolated and combined, a bimodal distribution is often observed \citep{furlanettooh,iliev14}. Simulations of characteristic bubble sizes at $z=6.6$, suggest that models with $R_b < 40$~$h^{-1}$cMpc are sufficient to cover all of the bubble distributions in the literature \citep{iliev14,mellema15,wyithe15}. Existing measurements from the Ly-$\alpha$ forest constrain the neutral fraction of the IGM at $z$=6.6 to be 20-50\%, corresponding to a temperature contrast of 5-12~mK according to the simulations of \citet{hutter17}. That work shows the expected brightness temperature contrast between ionised LAE regions and surrounding partially-ionised IGM as a function of neutral fraction and angular scale using using a set of reionisation simulations that couple a hydrodynamical GADGET-2 simulation \citep{springel05} with the radiative transfer code pCRASH \citep{partl11}. The results are not dependent on escape fraction, but suggest that significant temperature contrast is possible at $z>$6.5. It should also be noted that simulations are imperfect models for reality, and \citet{mason20} has explored the correspondence of observed LAEs in simulations with galaxy evolution, and demonstrated that small-scale mis-modelling can lead to unphysical results. Recent simulations by \citet{raste21} show a currently-favoured late reionization model retains structure at $z$=6.6, and the observational evidence for late reionization \citep{kulkarni19,nasir20,davies16,daloisio15,chardin17} ending near $z$=5.3 opens the possibility of the hydrogen neutral fraction being as high as 40\% at $z$=6.6 \citep{soltinsky21}. Moreover, the work of \citet{lidz21} with these new models suggests that bubbles of radius 10--20~cMpc are reasonable at this redshift, and simulations show that a lognormal distribution is a good model for the distribution of bubble sizes.  Topologically, \citet{friedrich11} found that the shape of ionised regions was most spherical at early and late times, with more complex morphology in between, while \citet{lee08} suggested that in late-reionization, the IGM transitions to an overlap phase of bubbles, but that this was not consistent with random neutral islands above $z$=6 and retained structure. For the purposes of this work, where the instrument limits us to an angular resolution of $\sim$2cMpc, the exact topology is not relevant.

The \textit{Silverrush Survey} \citep[][and accompanying \textit{Silverrush} papers]{ouchi18,shibuya17} observed five fields for LAEs using narrowband imaging around $z=5.7, 6.6$ with the Subaru Hypersuprime Cam (HSC). The fields are located mostly in the northern sky and chosen to overlap with fields used by other surveys, with the UltraDeep-SXDS field being the most southern (Declination -5 degrees). The survey catalogued 58 bright LAEs in a 1.4 degree $\times$ 1.4 degree field centred on the UD-SXDS field at $z$=6.58. \citet{kubota18} studied the detectability of ionised regions around LAEs via cross-correlation with the 21~cm signal observed with the MWA Phase I, finding that a detection was possible with 1,000 hours of observations. \citet{yoshiura21} used Generative Adversarial Networks to extract the 21~cm line profile from simulations around LAE galaxies in Subaru Hypersuprime Cam fields, estimating that MWA Phase II could detect this signal on scales larger than $k$ = 0.2$h$Mpc$^{-1}$ using all Subaru fields.

 At these low redshifts, when reionisation is almost complete, the expected 21~cm brightness temperature signal is weak, with standard models predicting 1--10~mK brightness temperature fluctuations over bubbles of tens of comoving Mpc scales \citep[0.1--0.5 degree angular scale, and 2--3~MHz spectral extent, at these redshifts][]{hutter17,raste21,koopmans15}. The MWA \citep{tingay13_mwasystem,wayth18,bowman13_mwascience} in its second phase is used in this work to measure the 21~cm signal. The weakness of the signal relative to the sensitivity of the Phase I MWA, and the MWA's relatively small filling factor in its core ($<$10 percent), place the detection and imaging of bubbles out of reach in less than many hundred hours of data (ignoring foregrounds).

However, the Phase II extended array configuration of the MWA has excellent $uv$-coverage over angular scales that are well-matched to recovering the bubbles. In addition, the availability of tens of LAEs (and therefore bubbles) improves the sensitivity, due to the increased effective signal. Unlike a regular cross-correlation study, in this work, we do not use direct HSC data, but rather the extracted parameters for these LAEs (precise redshifts and sky locations), to predict the 21~cm visibility measurements of the MWA, and perform a detection experiment. We compare the input model of the expected 21~cm signal with the residual radio images using a pre-whitening matched filter binary detection (signal-present/signal-absent). We compare the result of this detector with the theoretical performance.%, and that obtained from the same experiment undertaken with models where the distribution of LAE radii is varied.

The paper is structured as follows. In Section \ref{sec:methods} the prewhitening matched filter and signal template model are described, before the data and their preparation are detailed in Section \ref{sec:data}. Simulations of signal-present and signal-absent datasets are then constructed in Section \ref{sec:sims}, before the Results are presented and discussed in Sections \ref{sec:results} and \ref{sec:discussion}. Throughout, we use $H_0=100h$km/s/Mpc, and Planck 2015 cosmology \citep{planck15}. Vectors are denoted in bold-font, and covariance matrices are denoted by ${\bf K}$. Model images have no accent (${\bf S}$). Images observed through the instrument are denoted with a tilde ($\tilde{\bf S}$), and the expectation images are denoted with an overbar ($\bar{\bf S}$).

\section{Methods}\label{sec:methods}
%Postage stamp images in Stokes I at LAE locations will be extracted from MWA data, and stacked to increase the contrast between the ionised regions and surrounding neutral region. We will compare the result of this detector with the theoretical performance, and that obtained from the same experiment undertaken with models where the distribution of LAEs has been randomised. Clustered LAEs will be combined to one image to avoid noise correlation (Higuchi+ 2018).
We aim to perform a detection experiment, where a single statistic is extracted from the data and compared with expectations for a null detection and a signal-present detection. We do not use any HSC data; only basic parameters of the location (redshift, sky position) of LAEs, which are assumed to reside in the centres of HII bubbles. The \textit{Silverrush} survey targetted several key observational fields. Here we focus on the UD-SXDS field near RA: 02:18:00, Declination -05:00:00, described in \citet{shibuya17}, because it lies in the south in a relatively quiet region of the low-frequency radio sky. This one field is chosen initially for this work to best-align with the MWA's latitude. In future, the UD-COSMOS field at +2 degrees Declination may be used, but this field resides closer to the Galactic Plane, which we want to avoid due to the presence of diffuse emission. The other fields are too far north.

For the signal, we form integrated images around the locations of 58 LAEs at $z$=6.6. For the null experiment, we form model images based on a distribution of noise consistent with the data noise properties. We note that for an interferometer, where the total sky signal is not recovered (images are zero-mean), the null detector here represents a fully-ionised, or a fully-neutral and uniform, state, where there is no temperature structure imprinted on the temperature field by the 21~cm signal.

\subsection{LAE 21~cm model}
The IGM is modelled very simply; a uniformly-distributed partially-neutral IGM with a constant 21~cm brightness temperature, punctured by spherical comoving ionised bubbles, centred on the measured positions and redshifts of \textit{Silverrush} LAEs. The 21~cm brightness temperature of the IGM is set by the hydrogen spin temperature (gas kinetic temperature) and neutral fraction. The brightness temperature, $\Delta{T}_B$, is left as a parameter of our model, and we test a set of bubble sizes against our data. Three types of experiment are undertaken:
\begin{enumerate}
    \item The bubbles are assumed to be all of equal size, and the performance is assessed for an exactly matched filter
    \item The signal-present simulated data are modelled as having a Gaussian distribution of bubble sizes, and matched against a signal filter of the mean size
    \item The signal-present simulated data are modelled as having a log-normal distribution of bubble sizes, and matched against a signal filter of the mean size.
\end{enumerate}
The latter two cases form more realistic experiments, where there is a mismatch between the actual LAE size and the filter used. The performance of these detectors is expected to be degraded when there is a mismatch that is larger than the data angular resolution. Table \ref{table:bubbles} describes the bubble radii used for the models, and their corresponding angular and spectral size.
\begin{table}
\centering
\begin{tabular}{|c||c|c|c|c|}
\hline 
$z=6.6$ & $\bar{R}_b$ ($h^{-1}$cMpc) & $\sigma$ ($h^{-1}$cMpc) & $\Delta\theta$ (') & $\Delta\nu$ (MHz) \\ 
\hline \hline Matched
%& 5 & 2.9 & 800\\ 
& 10 & 0 & 5.8 & 1.6\\ 
& 15 & 0 & 8.7 & 2.4\\
& 20 & 0 & 11.6 & 3.2\\ 
\hline Gaussian
& 15 & 1 & 8.7 & 2.4\\
& 15 & 2 & 8.7 & 2.4\\
& 20 & 1 & 11.6 & 3.2\\
\hline Log-normal
& 10 & 3 & 8.7 & 2.4\\
& 15 & 2 & 8.7 & 2.4\\
\hline
\end{tabular}
\caption{Distributions of HII bubble sizes modelled in this work, in comoving and observed units.}\label{table:bubbles}
\end{table} 
For the larger bubbles, there will be overlap between the ionised regions, reducing the statistical advantage afforded by 58 individual LAEs. This complication also removes our ability to perform an analytic Fourier Transform to predict the measured visibilities for these data (circles transform to a $jinc$ function, a Bessel function of the first kind, normalised by the wavenumber) with a phase corresponding to their position with respect to the phase centre\footnote{In the case where the instrument primary beam effect is negligible, as it is here for a beam centred on a field of side length 1.4 degrees.}. However, given the overlap of bubbles, this model cannot be used, and we revert to a computational Fourier Transform to predict the model visibilities:
\begin{eqnarray}
    V(u,v;\nu) &=& \int_\Omega B(l,m) S_B(l,m) \exp{(-2\pi{i}(ul+vm))}\,\, dldm\nonumber\\
    &\simeq& \int_\Omega S_B(l,m) \exp{(-2\pi{i}(ul+vm))}\,\, dldm\nonumber\\
    &=& \displaystyle\sum_{j=1}^{58} \int S^j_B(l,m) \exp{(-2\pi{i}(ul+vm))}\,\, dldm
\end{eqnarray}
where $j$ denotes the $j$-th LAE, $S_B$ is the angular model for the bubble at frequency $\nu$, and we have ignored the contribution of the beam due to its insignificance at pointing centre. A flat-sky approximation is sufficient because the Subaru field is very small. The bubbles are modelled as simple hard-edged top-hat regions of zero brightness temperature embedded in a uniform background IGM. Interferometers measure temperature differentials, and so the underlying signal would resemble a zero temperature background punctured by bubbles observed in absorption located at position ($l_0,m_0$) and frequency $\nu_0$:
\begin{equation}
    S^j_B(l,m) = \begin{cases}
-S_B, \,\,\,\,(l-l_0)^2+(m-m_0)^2 \leq r_b(\nu)^2\\
0,\,\,\,\, \text{otherwise} \label{eqn:signal}
\end{cases}
\end{equation}
with $r_b(\nu)^2 = R_b^2 - \epsilon(\nu-\nu_0)^2$, $\epsilon$ a scaling of observed to cosmological units \citep{morales04}, and $S_B = 2k_BT_B/\lambda_j^2$ converts temperature to specific intensity for the model. The LAEs with spectroscopic redshifts are placed at their measured redshifts, however those with photometric redshifts are assumed to reside at $z=6.58$, with an error associated with this uncertainty that is unquantified in this work. In practise, the observed image is a convolution of Equation \ref{eqn:signal} with the system response function (the PSF or synthesized beam), which encodes the baseline sampling function and is constructed from the Fourier Transform of the weighted Fourier-space sampling function. We denote this instrument operator as $\mathcal{O}$, such that the observed mean signal is:
\begin{equation}
    \tilde{S}^j_B(l,m) = \mathcal{O}{S}^j_B(l,m).
\end{equation}
    
\begin{figure*}
\includegraphics[width=0.49\textwidth]{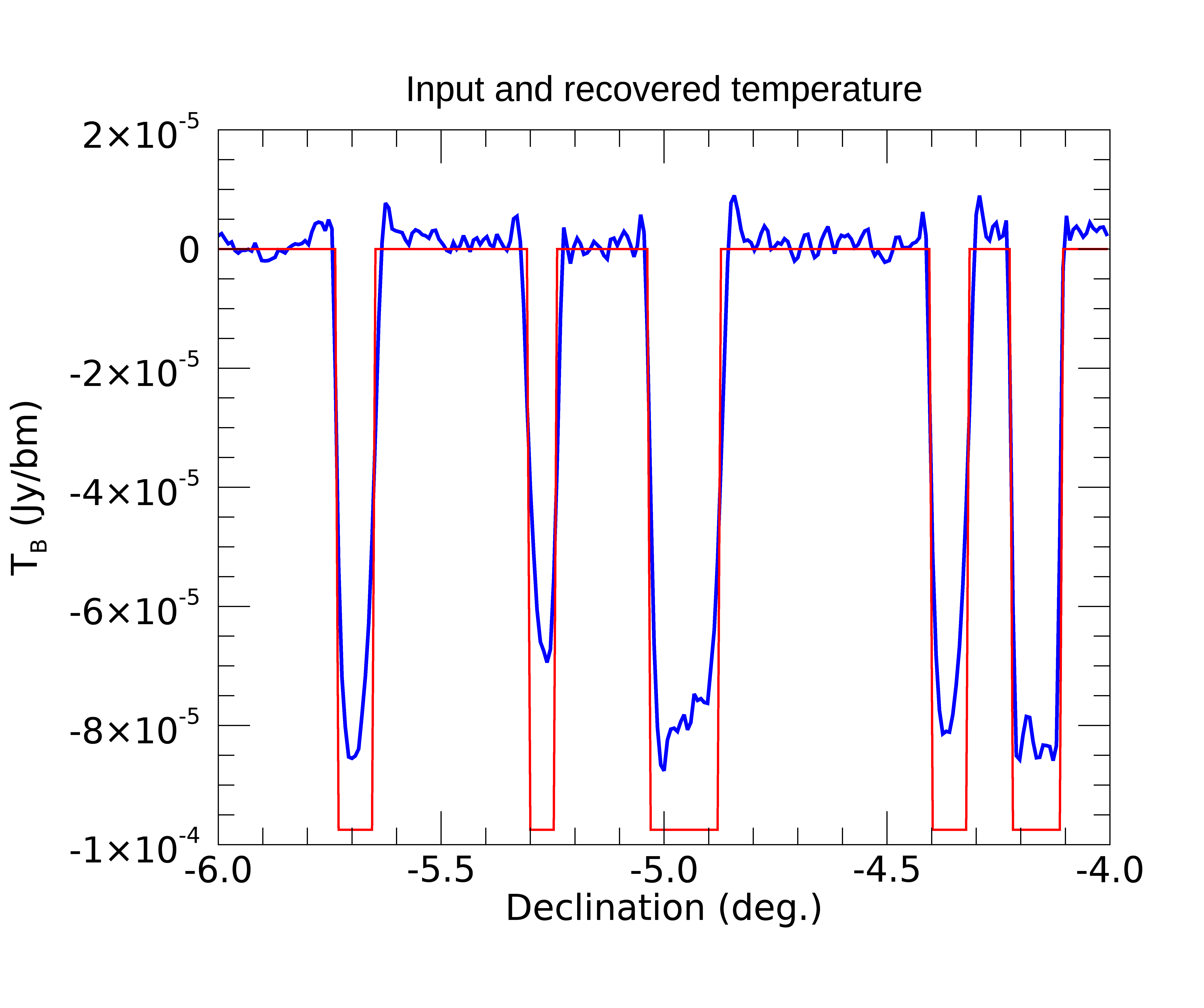}
\includegraphics[width=0.49\textwidth]{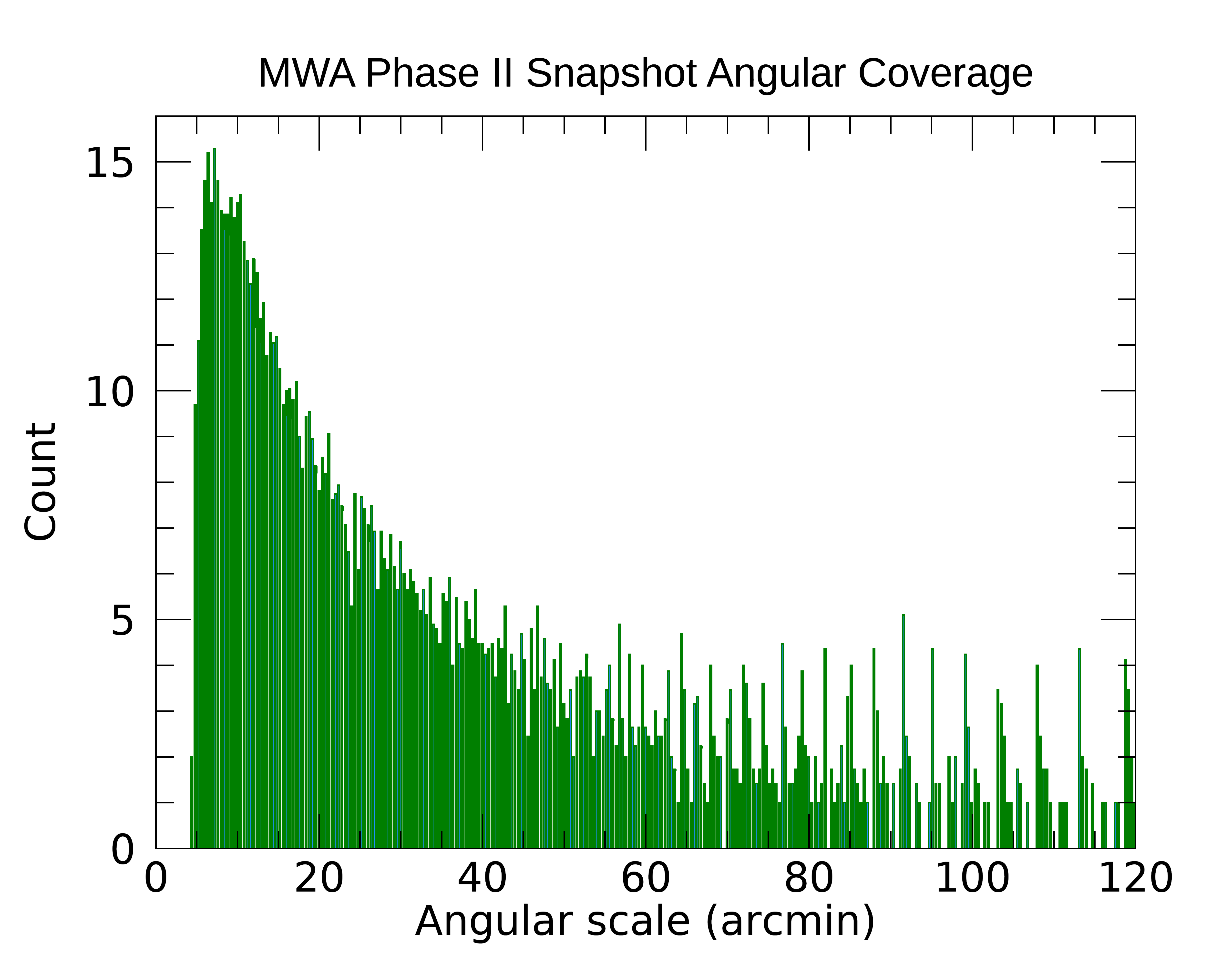}
\caption{(Left) Model (red) and recovered (blue) brightness temperature (K) field for 15$h^{-1}$cMpc bubbles with $T_B=1$~K for a single spectral channel (Jy/beam). The instrument sampling degrades the image, but the scales of relevance are well-recovered. Some loss of flux is visible within the bubble centres. (Right) Angular coverage of a meridian-pointed snapshot from MWA Phase II at 185~MHz, showing excellent coverage on scales of tens of arcminutes.}
\label{fig:model_bubbles}
\end{figure*}

A slice through the modelled brightness temperature distribution at $\nu=186$~MHz is shown in Figure \ref{fig:model_bubbles} (left) for bubbles of radius $R_b=15h^{-1}$cMpc, showing the input model (red) and expected measured signal (blue) after sampling with the MWA baseline distribution.
 The MWA's excellent $uv$-coverage allows us to recover the bubbles well, with minimal smearing and bowls from missing scales. Some loss of flux is visible within the bubbles due to missing scales, but a recovery of $\sim$90\% is found. The maximum amplitude of $T_B=1$~K corresponds to a trough of $\sim$0.1 mJy/beam for a single 80~kHz channel (see Section \ref{sec:obs}). The displayed signal is referenced to zero for the background, but an actual interferometer image would have zero mean due to the absence of the autocorrelation mode.
 
\subsection{Pre-whitening matched filter detector}
We aim to perform a detection experiment, and associate a level of confidence to the data being a statistical realisation of each model for the IGM. We use a Hotelling detector (pre-whitening matched filter detector with a statistical background), which provides optimal detection performance under Gaussian statistics with a signal-known-exactly, background-known-statistically (SKE-BKS) dataset \citep{barrettmyers,abbey97,kay98} (the optimal linear observer).

The detector matches the model signal template to a pre-whitened set of data, where the pre-whitening step accounts for background statistical uncertainty, statistical noise, and attempts to undo correlations between the data. It is used extensively in signal processing to test for the presence or absence of signals in data. We could choose to apply the detector equally to visibility data or image data; the Fourier Transform is a linear operation. In visibility space, statistical noise is uncorrelated, but residual background sources imprint structured noise. The Fourier Transform captures all of the signal in the telescope beam. In image space, the statistical noise is correlated on the scale of the imaging point-spread-function (PSF; the synthesized beam), computed from the baseline sampling function, and the residual background contains source confusion noise and source sidelobe noise which have similar properties. The noise covariance matrix in image space can be easily expressed with knowledge of the instrument baseline sampling, and can be inverted if the PSF is compact. Given the balance of arguments, we choose to apply the detector in image space, where its effects can be more readily visualised and its performance diagnosed.

We form two hypotheses: signal-present and signal-absent. The signal-present dataset, ${\bf x}_1$, contains 21~cm signal (${\bf S}_b$), correlated background noise ( mean ${\bf b}$, covariance ${\bf K}_b$) and correlated statistical noise (${\bf n}, {\bf K}_n$):
\begin{eqnarray}
    &{\bf x}_1& = {\bf \tilde{S}}_b + {\bf n} + {\bf b}\label{eqn:signpres}\\
    &{\bf x}_1& \sim \mathcal{N}\left( {\bf \tilde{S}}_b , {\bf K}_n+{\bf K}_b \right),
\end{eqnarray}
while the signal-absent dataset contains only correlated background noise and correlated statistical noise:
\begin{eqnarray}
    &{\bf x}_0& = {\bf n} + {\bf b}\label{eqn:signabs}\\
    &{\bf x}_0& \sim \mathcal{N}\left( 0 , {\bf K}_n+{\bf K}_b \right).
\end{eqnarray}
Here $\mathcal{N} \sim$ denotes that the data are distributed as Gaussian normal distributions.
The Hotelling observer constructs a linear template vector, ${\bf w}$,
\begin{equation}
    {\bf w} = ({\bf K}_b + {\bf K}_n)^{-1} \mathcal{O}\bar{\bf S}_b,
\end{equation}
where $\bar{\bf S}_b$ denotes the expected model signal template, corresponding to that described in Equation \ref{eqn:signal}, and we are operating on the expected signal \textit{after} it has passed through the instrument. The template is different for each experiment. Importantly, this includes the image data having zero-mean, because we do not recover the average sky temperature. For fields where the bubbles are large, the partially-neutral regions will be offset by the ionised regions. We will work in a regime where the radiometric noise dominates the source and sidelobe confusion ($\sim$1~mJy/bm), and omit these terms from the covariance matrix as they are absorbed into the measured noise properties of the data \citep{wayth18,franzen19}.

The template for each experiment is applied to the three dimensional MWA image data, ${\bf x}$, to form a scalar test statistic, $t$:
\begin{equation}
    t = {\bf w}^T {\bf x}. \label{eqn:teststat}
\end{equation}
The test statistic is used as the decision variable. The performance of the Hotelling observer can be constructed from the test statistic. The detection signal-to-noise ratio (SNR), $d$, is computed from the properties of the test statistic under the different hypotheses using the means and variances,
\begin{equation}
    d = \frac{\langle t_1 \rangle - \langle t_0 \rangle}{\sqrt{0.5({\rm var}(t_1)+{\rm var}(t_0))}}.
\end{equation}
For the Hotelling observer, the test statistic has the following properties under each hypothesis:
\begin{eqnarray}
    t_1 &\sim& \mathcal{N}(\bar{\bf S}_b^T\mathcal{O}^T{\bf K}^{-1} \mathcal{O}\bar{\bf S}_b,\bar{\bf S}_b^T\mathcal{O}^T{\bf K}^{-1} \mathcal{O}\bar{\bf S}_b)\\
    t_0 &\sim& \mathcal{N}(0,\bar{\bf S}_b^T\mathcal{O}^T{\bf K}^{-1} \mathcal{O}\bar{\bf S}_b),
\end{eqnarray}
where ${\bf K} = {\bf K}_b + {\bf K}_n$, yielding a detection SNR:
\begin{equation}
    d = \sqrt{\bar{\bf S}_b^T\mathcal{O}^T{\bf K}^{-1} \mathcal{O}\bar{\bf S}_b}.
\end{equation}
Intuitively, the SNR is maximised for model signals that are strong relative to the uncertainty. In this work, both the statistical (radiometric) noise and residual background source noise will play a role in setting the overall uncertainty, and the correlation of data on scales of the PSF will set the number of independent measurements we can obtain from the data.

With this framework, we can proceed with two important steps: (1) simulate signal-present and signal-absent datasets, accounting for the observational design and pre-processing of our data, and compute the expected detection performance; (2) apply the template to the data to measure a test statistic for a given input model (signal-present).%, and obtain real signal-absent datasets by shuffling the LAE positions and applying the same template. The latter is an important check to ensure that our data have the properties we expect given our statistical framework.

%We now describe the observations, analysis and dataset before describing the simulations.

\section{Observations and dataset}\label{sec:data}
\subsection{Observations}\label{sec:obs}
Data were observed in 2020 November (Extended), comprising 1,000 2-minute observations (33~hours) of the SXDS field (RA: 2h 18m, Dec.:-05.00 degrees) over six array pointing directions\footnote{Data were also obtained in 2017 August in Compact configuration, but these calibrated poorly and were omitted}. The data were observed over 24 1.28~MHz coarse channels in two frequency bands (181.755--191.995~MHz, 202.235--222.715~MHz) encompassing the two target redshifts for \textit{Silverrush} LAEs, $z$=5.7, 6.6. The eight lowest-frequency coarse channels were extracted to match the $z$=6.6 LAEs. The visibility data were observed at 1-second and 10~kHz resolution. Of the observations, we chose those from the meridian pointing (MWA gridpoint 25) and the four pointings contiguous to the meridian (gridpoints 29, 36, 37 and 57), to retain a well-behaved and understood instrumental beam. This resulted in a set of 874 observations (29 hours).

The extended array configuration is intended to have high angular resolution and good snapshot $uv$-coverage for survey programs and imaging. Its baselines range from 20~m to 5000~m with excellent snapshot coverage \citep{wayth18}. Figure \ref{fig:model_bubbles} (right) shows the angular coverage of a meridian snapshot from the Extended Array at 185~MHz for a Briggs weight of zero (used here) providing an intermediate weighting between natural and uniform. Structures with scales of 10s of arcminutes are well-recovered. The $uv$-plane is further improved by including the four off-meridian pointings.
%These visibilities are then gridded onto the $uv$-plane, as is performed by the CHIPS power spectrum software for processing regular MWA EoR data.

\subsection{Data reduction and processing}
The data were calibrated, peeled and imaged in several steps to produce the cleanest and deepest final images. The data were averaged to 8-seconds and 80~kHz resolution after flagging, direction-independent and -dependent calibration, and ionospheric quality assessment \citep{jordan17} using the MWA Real-Time System \citep{mitchell08}. This is the same data calibration as is applied to regular MWA EoR data, according to \citet{jacobs16} and \citet{trott16,trott20}. The 1000 brightest apparent sources were used to perform direction-independent calibration, and then 2000 were peeled using updated Jones matrices computed from the apparent positions of the five brightest calibrators. Tile 4 was flagged for all observations due to poor calibration solutions. For the $z=6.6$ range, each observation produced data for eight contiguous coarse channels, each with 16 80~kHz channels, spanning 10.24~MHz. Peeling of the 2000 apparent brightest sources for this field yields data with residual source flux densities of $<40$~mJy.

Following production of the UVFITS peeled and calibrated visibility data, data were converted to CASA measurement sets prior to imaging. We image each observation individually with WSClean \citep{offringa14,offringa-wsclean-2017,vandertol-2018} at low angular resolution and 80~kHz spectral resolution for further diagnostic assessment. No deconvolution was applied, and we only image Stokes I. The $z=6.6$ band includes frequencies of Perth and Geraldton digital TV (DTV) stations, and these have been detected in MWA data as reflections from satellites, aircraft etc. \citep{wilensky19}. We expect DTV-based RFI to be the primary cause for bad data. After inspection of the images, the list of 874 observations was reduced to 350 (11.7 hours). 
%Compact array data were found to be poorly calibrated due to the lack of $uv$-coverage and relatively poorer sky model at this location.
In general, sets of data spanning several minutes were found to be contaminated, and excised, while other periods of time were clean. Data excision was based on visual evidence for structure in the images, and excess image standard deviation across any coarse channel compared with the bulk of the distribution.

The data were then re-imaged with WSClean. We image with 3072 pixels on a side, each of 0.0075 degrees (27 arcsecond resolution -- oversampling PSF by a factor of three -- over a field of side length 23 degrees), with a Briggs weighting of zero to balance sidelobe suppression with noise level, and weighting by beam. We performed deconvolution with 1,500 iterations, which was required to remove the sidelobes from the small number of residual sources. The data were then averaged to form a single image cube of sides 3072$\times$3072$\times$128, spanning 23 degrees on a side and 10.24~MHz depth. Only the central 1.6 degrees is required for further analysis, representing a small subset of the field with little noise deviation. This smaller field is sufficient to encase the Subaru field-of-view, as well as a surrounding area that can be used for signal-absent datasets. The final dataset has an average rms noise level of 9--11 mJy/bm, which is a factor of $\simeq$1.5 times larger than the theoretical noise.

\begin{figure}
\includegraphics[width=0.49\textwidth]{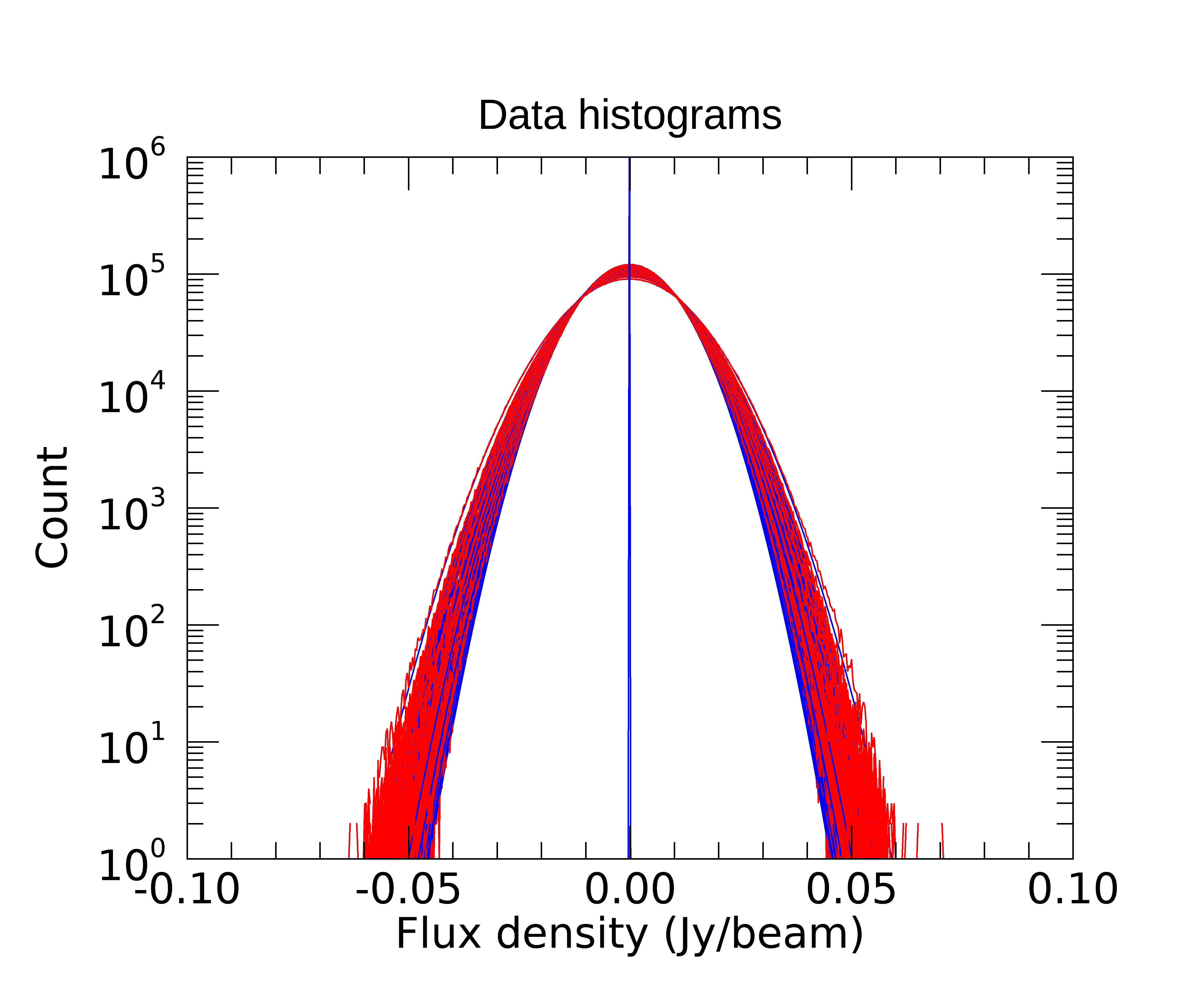}
\caption{(Histogram of image values for each channel (blue) and a Gaussian fit (red). The residual data cube is highly Gaussian. }
\label{fig:imagehist}
\end{figure}

\begin{figure}
\includegraphics[width=0.45\textwidth]{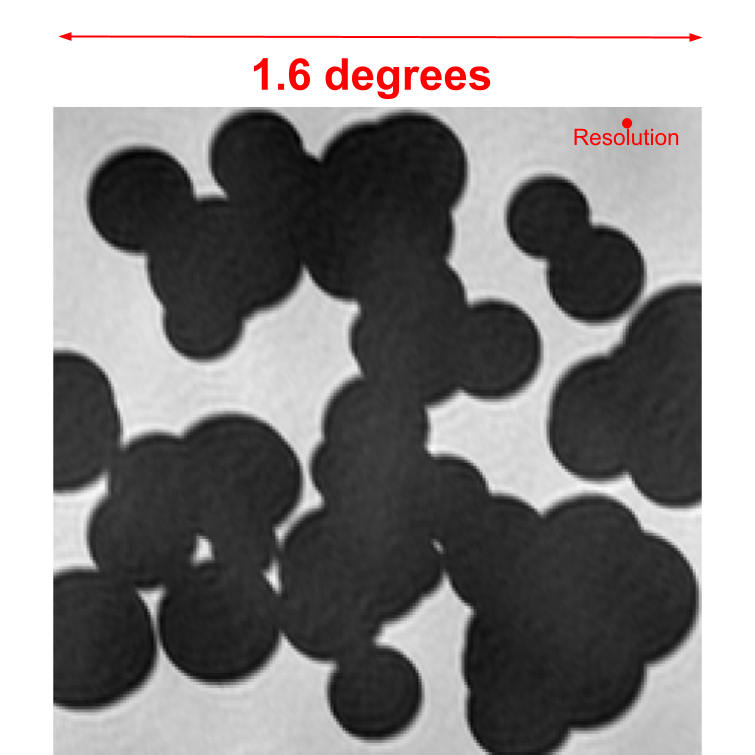}
\caption{Example expected signal from ionised regions with a mean radius of 15~$h^{-1}$cMpc and standard deviation of 2~$h^{-1}$cMpc, after instrument sampling, at 185~MHz. The angular size is shown, as well as an indicative MWA resolution element.}
\label{fig:ds9}
\end{figure}

Figure \ref{fig:imagehist} shows the histogram of residual image values for each channel (red) and a Gaussian fit (blue). The residual data cube is highly Gaussian, with small tails of negative and positive pixels. The positive pixels generally coincide with weak sources that remain in the data, while negative pixels coincide with the peeling residuals around bright sources.
A slice through the expected signal template cube is displayed in Figure \ref{fig:ds9} at 185~MHz, matching the size of the UD-SXDS field from Subaru observations. The field shows bubbles of mean radius 15~$h^{-1}$cMpc, and standard deviation of 2~$h^{-1}$cMpc.

%\subsection{Post-calibration spectral subtraction}
%The bandwidth of the data is intended to allow for a sufficient lever arm to perform a spectral fitting of a smooth function for further foreground removal. The ionised regions we are probing have bandwidths smaller than 3.5~MHz, and we can therefore use the full 10.24~MHz observed band to remove some of the residual foreground contamination. [Do we add this back into the model to get the K constraints\?]

\section{Simulations}\label{sec:sims}
We begin by forming the covariance matrix of the data. Each channel is treated individually, and so we need to construct the relationship between each pixel in the image. The covariance matrix is formed by describing the PSF convolution matrix as a vector, ${\bf B}$ and using a diagonal matrix ${\bf D}$ to form ${\bf K} = {\bf BDB^T}$. The noise is heteroscedastic across each slice, consistent with the measured rms. The 2D images are collapsed to one dimension by taking a lexicographical ordering of the pixels. We construct a two-dimensional covariance matrix from the one-dimensional data.

The data and images are placed into a consistent coordinate system of pixel scale 0.0075 degrees and 213 pixels on each side, resulting in images of size 1.6 degrees $\times$ 1.6 degrees to encompass the Subaru field-of-view. The central 120 spectral channels are selected to match the Subaru narrowband filter. Of these, 14 are missing and omitted from the analysis, leaving 106 viable channels. Each channel is independent, and so the full covariance is block diagonal. As such, each channel is treated separately and the test statistic summed over frequency. In one dimension for a single channel, this results in data of length 213$\times$213$=$45,369 pixels, and we therefore construct a 45,369$\times$45,369 pixel covariance matrix. Fortunately, the imaging PSF is compact, and is well-fitted by a rotated elliptical Gaussian with semi-major and -minor axes of 33" and 26" respectively. We use a 13$\times$13 pixel (5.9'$\times$5.9') cut-out of the PSF to construct the covariance matrix. The covariance is highly-diagonal and very sparse, allowing for accurate solution of the linear equation;
\begin{equation}
    t = {\bf w}^T {\bf x} = \bar{\bf s}^T {\bf K}^{-1} {\bf x}.
\end{equation}
We use Cholesky decomposition to solve the linear system of equations (we do not invert the covariance matrix separately) through the IDL functions \textsf{LA\_CHOLDC} and \textsf{LA\_CHOLSOL}.

\begin{figure}
\includegraphics[width=0.48\textwidth]{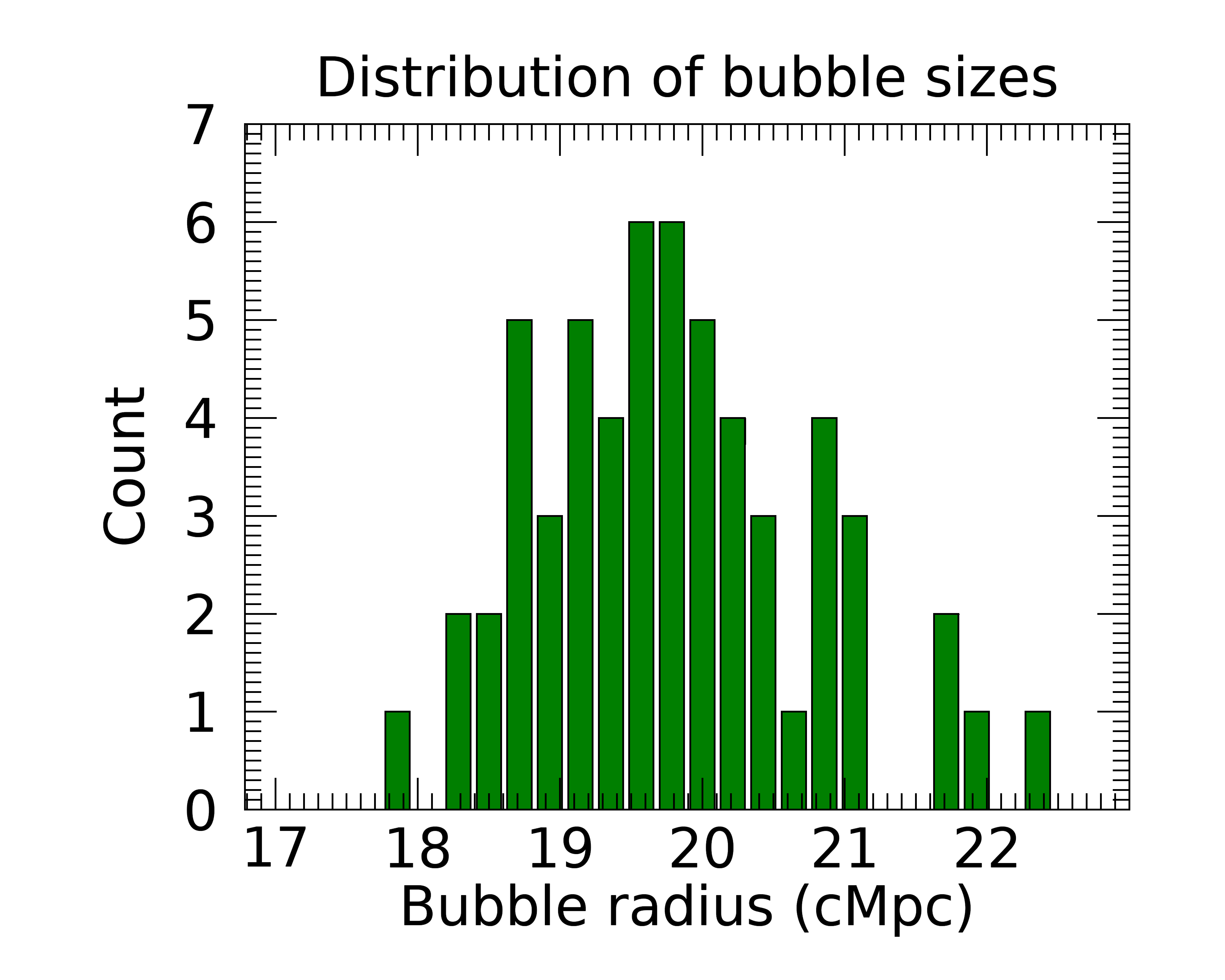}
\caption{Gaussian distribution of sizes for one realisation with a mean bubble size of 20~$h^{-1}$cMpc, and standard deviation of 1~$h^{-1}$cMpc.}
\label{fig:bubble_dist}
\end{figure}

Simulations of signal-present and signal-absent datasets are used to determine the level of significance for the measured test statistic from the data. In all cases, the test signal (matched filter) uses uniform bubble sizes because there is no \textit{a priori} knowledge of the size of an individual bubble. As described earlier, three types of experiment are then conducted; (1) uniform bubbles (an exact matched filter), (2) Gaussian-distributed bubbles with a matched mean size, (3) Lognormally-distributed bubbles with a matched mean size.
%After performing the experiment for the simplistic case of uniformly-sized bubbles, complexity is added via a distribution of sizes. The bubble sizes are Gaussian-distributed with a mean of 15~cMpc or 20~cMpc, and a standard deviation of 1~cMpc and 2~cMpc. The simulated datasets with non-uniform bubble sizes are matched to an expected signal with the mean bubble size, to emulate the case where a simple matched signal is applied to real data. 
I.e., a mismatch between the data and expected signal is modelled in the simulation to test the performance when the exact bubble size distribution is unknown. The performance is expected to be degraded, but not significantly for deviations in bubble size that are comparable to the instrument resolution (2~$h^{-1}$cMpc). Figure \ref{fig:bubble_dist} shows the distribution of sizes for one realisation with a mean bubble size of 20~$h^{-1}$cMpc, and standard deviation of 1~$h^{-1}$cMpc (Gaussian-distributed).

Signal-present and -absent datasets are constructed according to Equations \ref{eqn:signpres} and \ref{eqn:signabs}. The model temperature distribution for each 80~kHz spectral channel is Fourier Transformed to the $uv$-plane, sampled according to the instrument baseline distribution for the observations, and transformed back to image space. The brightness temperature is converted to measured units of Jansky per beam using the beam size extracted from the WSClean imaging. Finally, the  simulated signal is mean-subtracted to match that for a measured field. This forms the matched filter template expected signal, ${\bf \bar{S}}_b$.

The signal-present datasets with uniform bubble size are constructed by adding the expected signal to zero-mean Gaussian-distributed noise matching that measured for the data, in each spectral channel. The signal-present datasets for non-uniform bubble size are computed independently, and Gaussian-distributed noise added to them. The signal-absent datasets contain only Gaussian-distributed noise. There are 15,000 realisations simulated for each of the signal-present and -absent datasets.

Note that the signal-present dataset is equivalent to the signal-absent dataset when (1) reionisation is complete (late times), or (2) reionisation has not commenced, and the 21~cm temperature field is uniform over the field (early times). A field where half of the volume is ionised will provide the maximal signal. Within the UD-SXDS field, the volume filling factors for the 10~$h^{-1}$cMpc, 15~$h^{-1}$cMpc, and 20~$h^{-1}$cMpc bubbles are 23\%, 53\%, and 80\%, placing the 15~$h^{-1}$cMpc bubbles close to the maximal size.

\begin{figure}
\includegraphics[width=0.45\textwidth]{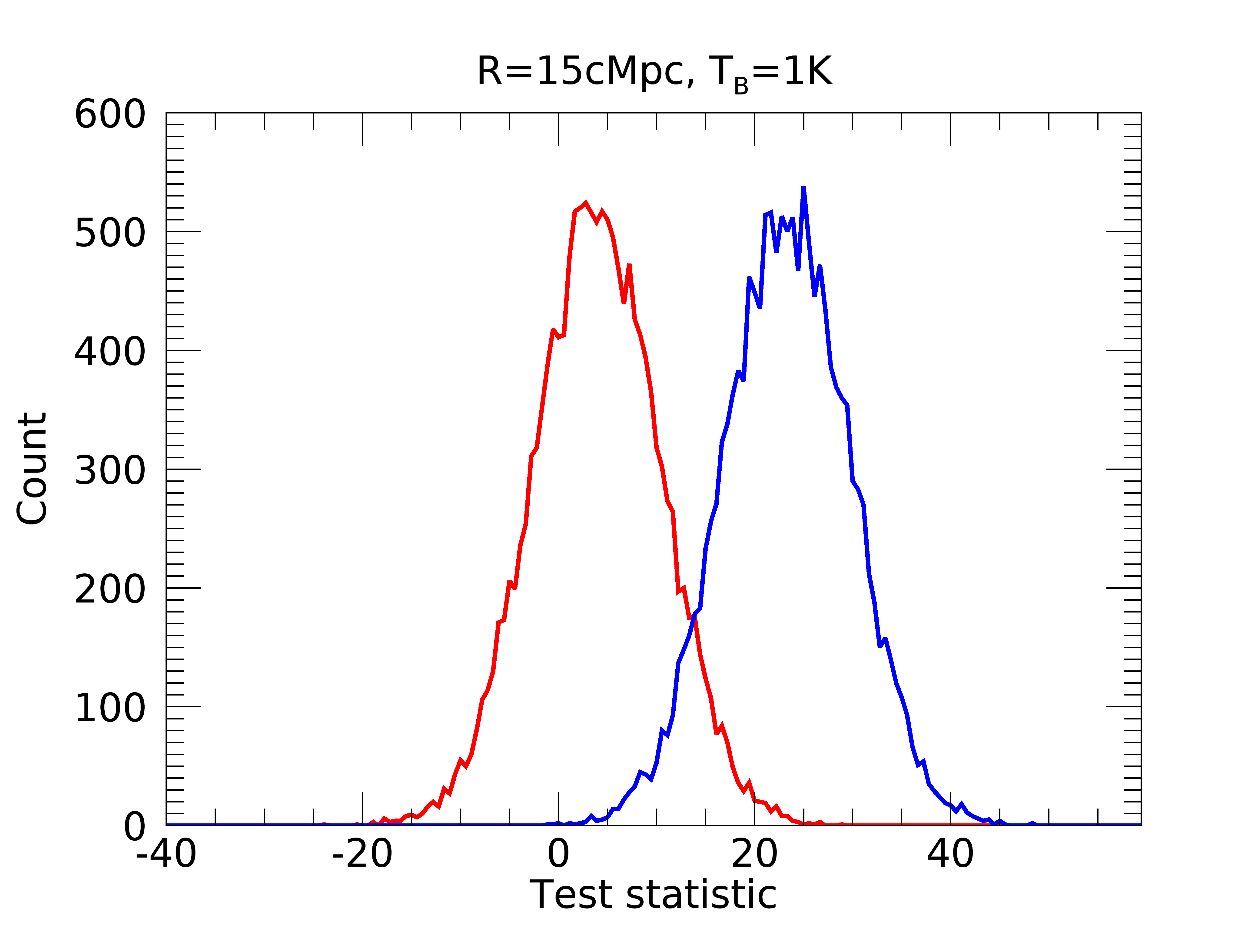}
\caption{Distribution of test statistics for realisations of signal-present (blue) and signal-absent (red) datasets matched to a model signal with bubble size 15~$h^{-1}$cMpc and a 1~K IGM brightness temperature.}
\label{fig:signalpresabs}
\end{figure}

Figure \ref{fig:signalpresabs} displays histograms of the test statistic, $t$ (Equation \ref{eqn:teststat}), for a 1~K IGM brightness temperature punctured by bubbles of radius 15~$h^{-1}$cMpc.
The width of the distributions is defined by the noise level in each residual dataset, and the number of independent measurements in the datacube. As the brightness temperature of the medium is reduced, the separation of the two histograms will reduce. These simulated test statistics will be used to set limits on the temperature for a given bubble size.

\section{Results}\label{sec:results}
\begin{figure}
\includegraphics[width=0.49\textwidth]{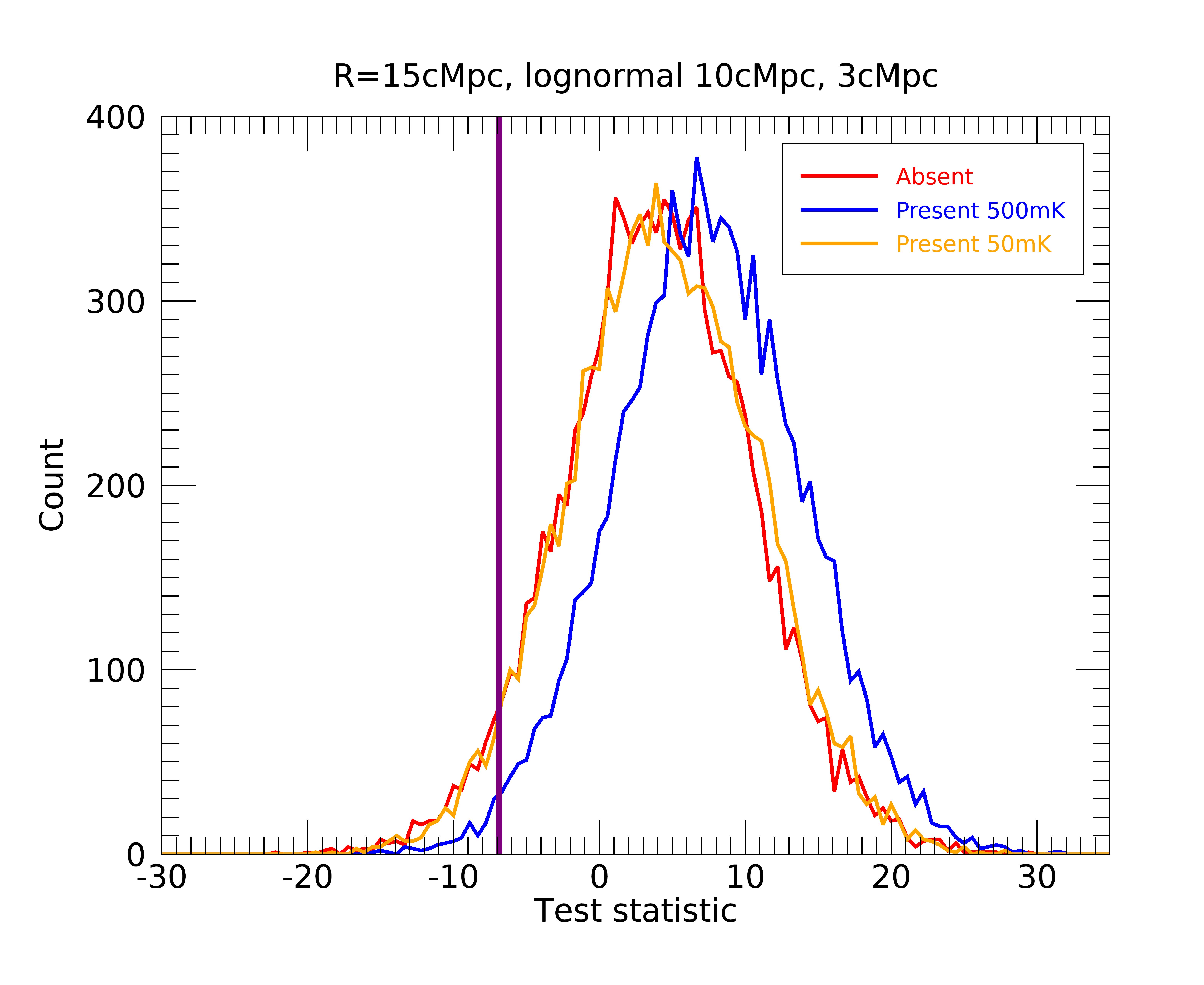}
\caption{Signal-absent (red), signal-present 50mK (orange), and signal-present 500mK (blue) distributions of test statistics for the lognormally-distributed data matched to a uniform 15$h^{-1}$cMpc signal template. Also shown is the measured test statistic, $t=-6.89$, from the data (vertical purple line).}
\label{fig:result}
\end{figure}

Three matched filter signals are applied to the data, one for each bubble radius (10, 15, 20~$h^{-1}$cMpc). The result test statistics for applying these signals to the data are compared for each experiment. Figure \ref{fig:result} shows the signal-absent (red), signal-present 50mK (orange), and signal-present 500mK (blue) distributions of test statistics for the lognormally-distributed data matched to a uniform 15$h^{-1}$cMpc signal template. Also shown is the measured test statistic, $t=-6.89$, from the data (vertical purple line).
Note that the signal-absent test statistic distributions are skewed positive, while both the data and template are zero-mean. This is partially due to the effect of the prewhitening filter, and the off-diagonal (pixel correlation) terms. \footnote{The covariance matrix is constructed using an estimate for the system PSF, and there is therefore a small mis-match between the true data covariance and the pre-whitening filter. This has the desired effect of accounting for the reduction in independent measurements, but does not completely undo the data correlations.} The exact same prewhitening filter is applied to all experiments, and the signal-absent and signal-present datasets, and so the actual values of the test statistic are not relevant; only their distributions, and the results are not biased. \footnote{Because the data and covariance matrix are not independent (by construction as the covariance matrix is designed to capture the correlation in the image), a signal-absent noise-only dataset still contains correlations due to its convolution with the system PSF to generate realistic signal-absent data.}
Table \ref{table:results} lists the measured test statistics for the data matched to each uniform-bubble signal template, and the IGM brightness temperature limits that are excluded at the 68\%, 95\% and 99\% levels of confidence. These limits are determined by comparing the measured test statistics to the signal-present simulations for each of the three experiments (i.e., where the signal-present test data have uniform, Gaussian-distributed, and lognormally-distributed bubble sizes).
\begin{table}
\centering
\begin{tabular}{|c||c|c|c|c|c|c|c|c|}
\hline 
$z=6.6$ & $\bar{R}_b$ & $\sigma$ & ${\bar{S}}_b$ & $t$ & 68\% & 95\% & 99\% \\ 
\hline \hline Matched
%& 5 & 2.9 & 800\\ 
& 10 & 0 & 10 & 4.04 & 350~mK & 840~mK & $<$1~K \\ 
& 15 & 0 & 15 & -6.89 & 0~mK & 30~mK & 200~mK \\
& 20 & 0 & 20 & -7.30 & 0~mK & 50~mK & 200~mK\\ 
\hline Gaussian
& 15 & 1 & 15 & -6.89 & 0~mK & 30~mK & 200~mK \\
& 15 & 2 & 15 & -6.89 & 0~mK & 30~mK & 200~mK \\
& 20 & 1 & 20 & -7.30 & 0~mK & 30~mK & 250~mK \\
\hline Lognorm
& 10 & 3 & 15 & -6.89 & 0~mK & 50~mK & 500~mK \\
& 15 & 2 & 15 & -6.89 & 0~mK & 30~mK & 200~mK \\
\hline
\end{tabular}
\caption{Measured test statistics for the data matched to each uniform-bubble signal, and the IGM brightness temperature limits that are excluded at the 68\%, 95\% and 99\% levels of confidence. These limits are determined by comparing the measured test statistics to the signal-present simulations for each of the three experiments. The first three numerical columns give the simulated signal-present bubble size and standard deviation ($\bar{R}_b, \sigma$), and the matched signal template, $\bar{S}_b$.}\label{table:results}
\end{table}

For all of the experiments, the measured test statistics are consistent with noise for the 15$h^{-1}$cMpc and 20$h^{-1}$cMpc bubble sizes. As the temperature of the IGM is increased from zero, the signal-present histograms move to the right, and the measured test statistics become inconsistent with these models at some level of significance. Due to the small values of $t$ measured for the 15$h^{-1}$cMpc and 20$h^{-1}$cMpc signal templates (both negative), an IGM temperature of tens of mK is inconsistent with the data at 95\% confidence, and 200--500mK at 99\% confidence, depending on the model, noting that a noise-only model is equally-favoured. The extreme sensitivity to the IGM temperature is due to the measured $t$ falling at a low (but non-zero) likelihood value for noise-only data, however an increase in the IGM temperature would make the noise-only model favoured strongly over the signal-present model, implying that there is sensitivity to non-zero signal. The data matched to a 10$h^{-1}$cMpc signal template yields a larger value of $t=4.04$, providing much weaker constraints for this bubble size.

The results are not significantly degraded when we move from the idealised case of a signal template perfectly-matched to the data, to a more realistic Gaussian or lognormal distribution of bubble sizes. In these experiments, the limits are unchanged at the 95\% level, but somewhat degraded at 99\%, reflecting that there is mismatch. This is consistent with these models having size variations that are comparable to the instrument resolution, and retaining the same central locations for the bubbles (i.e., aligned with the LAE locations from \textit{Silverrush}).

\section{Discussion}\label{sec:discussion}
Increasing the amount of clean data will reduce the rms noise, narrowing the distributions of signal-present and signal-absent test statistics. Those tests would have increased discrimination power and the ability to place more stringent limits, or detect, the IGM temperature. In that case, it would be worth doing a second round of tests where the bubble locations were randomised to separate data systematics from stochastic variations (i.e., if the data have residual non-cosmological structure, otherwise unaccounted for in the model). This would provide a robust test for the detection of ionised and non-ionised regions. For this work, where the noise level is only sufficient to place weak constraints, it is not necessary to go to such lengths.

The work presented here still suffers from having to assume something about the distribution of bubble sizes, and the brightness temperature spatial distribution. The simple model assuming a single temperature across the partially-neutral regions is simplistic, and really what we are probing is a volume-averaged brightness temperature in this work. A lot of detection power is afforded by the simple assumption that the LAEs reside at bubble centres, but strong deviations of the template from the data results in degraded detection performance. The limited resolution of the MWA helps in this respect, allowing models with offsets of $\sim$2-3$h^{-1}$cMpc to be absorbed into the loss of signal from the instrumental angular resolution, thereby hiding more complex structure to the ionisation fronts. This means that if reality is represented by the lognormal distributions observed in simulations, the simple model is still sufficient to perform a detection experiment. A large-scale change in the topology of reionisation (e.g., a significant bimodal size distribution or qualitatively different evolution) will lead to degraded results as the matched signal filter deviates more substantially from the model. However, given the consistency of different simulations with the models chosen in this work, this is encouraging for future experiments of this kind. With increased data (reduced noise), a full Bayesian analysis can be performed to marginalise over the free parameters (bubble mean size and standard deviation, and IGM temperature) to produce a robust detector.% We should also mention that with only one field used in this work, cosmic variance has not been quantified, and the results may be the result of the 'look-elsewhere effect', where we may be viewing an unlikely volume of the Universe.

The results of this work can be placed in the context of the expectations for this level of noise. The measured noise level of $\sim$10 mJy/bm, translates to a surface brightness rms temperature of 80~K in each independent element for each of the 106 useable channels. From a pure noise perspective, there are $\sim$~1 million independent measurements in the data cube, yielding an expected temperature resolution of $\sim$80~mK. This matches with the width of the test statistic distributions, and ultimately limits the ability to place strong constraints on the IGM temperature. Nonetheless, the technique shows promise for a detection with a larger dataset. In this work, more than half of the data were removed due to poor calibration and contamination. The residual dataset is very clean, and well-behaved, showing Gaussian-distributed noise statistics. With four times as much data, the noise level would provide stronger constraints.

Ultimately, this work would benefit from a larger field of detected LAEs. The \textit{Silverrush} survey encompassed 21 square degrees, an improvement by a factor of ten compared with the single UD-SXDS field. It also included $z=5.7$ LAEs. In the late-reionization models currently favoured by Ly-$\alpha$ forest measurements, this redshift bin may contribute useful information. At $z$=6.6, the larger area of the full survey would improve the detectability (SNR) by $\sqrt{10}$. Alternately, a larger survey around the same UD-SXDS field would be advantageous for matching to the MWA field-of-view and latitude, where a 21 square degree field would still easily reside within the instrument's main lobe. In future, it is the survey area of these optical LAE programs that will be the limiting factor for experiments with the MWA and SKA.

\section{Conclusions}

In this work, a pre-whitening matched filter detector of ionised regions around detected LAE galaxies is applied to 21~cm data to constrain the brightness temperature of the IGM at $z$=6.6. This is the first time such an experiment has been attempted. The 12 hours of clean MWA data spanning $\sim$10~MHz and 1.6$\times$1.6 square degrees had a residual rms noise of 10 mJy/bm after calibration and point source peeling. Point and extended sources are used for calibration and peeling, but there is no diffuse model implemented because it is not currently part of the sky model. Galactic Synchrotron is expected to be coherent over scales larger than the Subaru field-of-view, but some will be on the same scale. This field is away from the Galactic Plane sufficiently to do this work, and is one motivation for choosing this \textit{Silverrush} field. A 3D signal template was created from the known sky and redshift locations of LAEs from the \textit{Silverrush} survey for bubbles of different radius, and the detector's performance compared with the expectations from simulated signal-present and signal-absent datasets. Assuming the bubble sizes are a priori known and of uniform radius of 15$h^{-1}$cMpc, an IGM with a brightness temperature exceeding 30 mK is ruled-out at 95\% confidence, and 200 mK at 99\% confidence. For a more realistic model where a uniform size signal template is matched to data with a lognormal distribution of bubble sizes, the performance is slightly degraded with an IGM temperature of 50 mK ruled-out at 95\% confidence. These results are consistent with theoretical expectations for the brightness temperature of the IGM at this redshift of 1--10mK. Future work with more data can employ a fully-Bayesian framework to perform a detection experiment that marginalises over key parameters, such as IGM temperature, mean bubble radius, and bubble radius standard deviation.

\section*{Acknowledgements}
We thank the referee for the many comments that have clarified the methodology and improved the exposition, in particular the discussion about the bias in the test statistic.
We would like to thank Marcin Sokolowski and Andrew Williams for help with scheduling the observations.
This research was partly supported by the Australian Research Council Centre of Excellence for All Sky Astrophysics in 3 Dimensions (ASTRO 3D), through project number CE170100013. CMT is supported by an ARC Future Fellowship under grant FT180100321.
The International Centre for Radio Astronomy Research (ICRAR) is a Joint Venture of Curtin University and The University of Western Australia, funded by the Western Australian State government. KT is partially supported by JSPS KAKENHI Grant Numbers JP15H05896, JP16H05999 and JP17H01110, and Bilateral Joint Research Projects of JSPS.
PD and AH acknowledge support from the European Research Council's starting grant ERC StG-717001 (\textit{DELPHI}). PD acknowledges support from the NWO grant 016.VIDI.189.162 (\textit{ODIN}) and the European Commission's and University of Groningen's CO-FUND Rosalind Franklin program.
The MWA Phase II upgrade project was supported by Australian Research Council LIEF grant LE160100031 and the Dunlap Institute for Astronomy and Astrophysics at the University of Toronto.
This scientific work makes use of the Murchison Radio-astronomy Observatory, operated by CSIRO. We acknowledge the Wajarri Yamatji people as the traditional owners of the Observatory site. Support for the operation of the MWA is provided by the Australian Government (NCRIS), under a contract to Curtin University administered by Astronomy Australia Limited. We acknowledge the Pawsey Supercomputing Centre which is supported by the Western Australian and Australian Governments.

%%%%%%%%%%%%%%%%%%%%%%%%%%%%%%%%%%%%%%%%%%%%%%%%%%
\section*{Data Availability}

The Murchison Widefield Array data used in this work are proprietary until July 2022, when they will be publicly-available via the ASVO MWA data page. The LAE locations used in this work can be obtained from the \textit{Silverrush} survey database.



\bibliographystyle{mnras}
\bibliography{manuscript} % if your bibtex file is called example.bib

%\begin{thebibliography}{99}
%\end{thebibliography}

%%%%%%%%%%%%%%%%%%%%%%%%%%%%%%%%%%%%%%%%%%%%%%%%%%

%%%%%%%%%%%%%%%%% APPENDICES %%%%%%%%%%%%%%%%%%%%%

% Don't change these lines
\bsp	% typesetting comment
\label{lastpage}
\end{document}